\ifdefined\pdfoutput\pdfoutput=1\fi
\newif\ifquickcompile
\ifdefined\quickcompile
	\quickcompiletrue
\else
	\quickcompilefalse
\fi
\ifquickcompile
	\PassOptionsToPackage{draft}{graphicx}
	\documentclass[12pt,draft]{article}
\else
	\documentclass[12pt]{article}
\fi
\usepackage{amsmath}
\usepackage{graphicx}
\ifquickcompile
	\renewcommand{\includegraphics}[2][]{%
		\fbox{%
			\begin{minipage}[c][0.12\textheight][c]{0.85\linewidth}
				\centering\footnotesize Figure omitted in quick compile\\
				\texttt{\detokenize{#2}}
			\end{minipage}%
		}%
	}
\fi
\usepackage{enumerate}
\usepackage{natbib}
\usepackage{url}
\usepackage{enumitem}

\usepackage{comment}
\usepackage[dvipsnames]{xcolor}
\pagecolor{white}
\usepackage{ulem}
\usepackage{cancel}
\usepackage{cancel}
\usepackage{multirow}
\usepackage{diagbox}
\usepackage{subfigure}
\usepackage{amsfonts}
\usepackage{float}
\usepackage{epsfig}
\usepackage{mathrsfs}
\usepackage{array}
\usepackage{verbatim}
\usepackage{cases}
\usepackage{color}

\newcommand{\blind}{0}

\newcommand{\PreserveBackslash}[1]{\let\temp=\\#1\let\\=\temp}
\newcolumntype{C}[1]{>{\PreserveBackslash\centering}p{#1}}

\newcommand{\vTheta}{\boldsymbol{\Theta}}
\newcommand{\bx}{\boldsymbol{x}}

\usepackage[letterpaper]{geometry}
\def\spacingset#1{\renewcommand{\baselinestretch}%
	{#1}\small\normalsize} \spacingset{1}

\usepackage[colorlinks,bookmarksopen,bookmarksnumbered,citecolor=blue,urlcolor=blue,linkcolor=blue]{hyperref}
\usepackage{natbib}

\usepackage{amsthm,amssymb,amsmath}

\newtheorem{theorem}{Theorem}

\newtheorem{lemma}{Lemma}
\newtheorem{corollary}{Corollary}

\newtheorem{definition}{Definition}
\newtheorem{assumption}{Assumption}

\usepackage[ruled, linesnumbered]{algorithm2e}

\usepackage{grffile,verbatim}
\usepackage{enumitem}
\usepackage{array,multirow}

\usepackage{graphicx}
\usepackage{caption}

\usepackage{color}
\usepackage{comment}

\usepackage{prettyref,soul}
\newrefformat{corollary}{Corollary~\ref{#1}}
\newrefformat{lemma}{Lemma~\ref{#1}}
\newrefformat{theorem}{Theorem~\ref{#1}}
\newrefformat{assumption}{Assumption~\ref{#1}}
\newrefformat{condition}{Condition~\ref{#1}}

\newcommand{\argmin}{\ensuremath{\operatornamewithlimits{argmin}}}

\def\trans{^{\rm T}}

\newcommand{\bbR}{\mathbb{R}}

\newcommand{\bbE}{\mathbb{E}}
\newcommand{\bbP}{\mathbb{P}}

\newcommand{\sG}{\mathcal{G}}
\newcommand{\sI}{\mathcal{I}}
\newcommand{\sA}{\mathcal{A}}

\newcommand{\sF}{\mathcal{F}}
\newcommand{\sH}{\mathcal{H}}
\newcommand{\sL}{\mathcal{L}}

\newcommand{\sP}{\mathcal{P}}

\newcommand{\va}{\mathbf{a}}

\newcommand{\ve}{\mathbf{e}}

\newcommand{\vx}{\mathbf{x}}

\newcommand{\valpha}{\boldsymbol{\alpha}}

\newcommand{\vbeta}{\boldsymbol{\beta}}
\newcommand{\vtheta}{\boldsymbol{\theta}}

\newcommand{\vgamma}{\boldsymbol{\gamma}}

\newcommand{\vtau}{\boldsymbol{\tau}}
\newcommand{\vzero}{\mathbf{0}}

\newcommand{\mB}{\mathbf{B}}
\newcommand{\mC}{\mathbf{C}}

\newcommand{\mS}{\mathbf{S}}

\newcommand{\mV}{\mathbf{V}}
\newcommand{\mW}{\mathbf{W}}
\newcommand{\mX}{\mathbf{X}}

\usepackage{relsize}

\def\spacingset#1{\renewcommand{\baselinestretch}
	{#1}\small\normalsize} \spacingset{1}
\allowdisplaybreaks

\begin{document}

	\def\spacingset#1{\renewcommand{\baselinestretch}%
		{#1}\small\normalsize} \spacingset{1}

	\newcommand{\blu}{\color{blue}}
	\newcommand{\gre}{\color{ForestGreen}}

	\if0\blind
{
	\title{\bf A Tensor Neural Network  for Interpretable and Predictive Representations}
	
	\author{Shuoli Chen$^{1}$, Zhiyao Tan$^{1}$, Li Liu$^{2}$ \  and  Huazhen Lin$^{1,3}$\thanks{Corresponding author. Email address: \emph{linhz@swufe.edu.cn}. } 
		\vspace{.5cm}\\
		$^1$ Center of Statistical Research and School of Statistics and Data Science,\\
		Southwestern University of Finance and Economics, Chengdu, China\\
		$^2$        School of Mathematics and Statistics,\\
		Wuhan University, Wuhan, China\\
		$^3$ New Cornerstone Science Laboratory, Chengdu, China }
	\maketitle
} \fi

	\if1\blind
	{
		\bigskip
		\bigskip
		\bigskip
		\begin{center}
			{\LARGE\bf A Tensor Neural Network  for \\
				Interpretable and Predictive Representations}
		\end{center}
		\medskip
	} \fi

	\bigskip
\spacingset{1}
	\begin{abstract}

Tensor regression is challenged by the exponential dimensionality of tensor-valued covariates and their complex multi-mode dependence. Existing methods therefore often rely on simple models, such as linear tensor regression and tensor additive models, which are computationally tractable but impose restrictive structural assumptions and may limit predictive accuracy. To address these limitations, we propose a tensor neural network that learns low-dimensional, interpretable, and prediction-sufficient representations while preserving the multi-mode structure of tensor data. The model captures nonlinear effects for accurate response prediction while retaining the interpretability of classical linear tensor regression. Theoretically, we establish selection consistency for the minimal representation dimension \(K^*\) necessary for prediction and show that the estimator attains a representation-adaptive non-asymptotic error rate. Specifically, the rate improves from the ambient-dimensional order \(n^{-\frac{\beta}{\mathcal{D}+2\beta}}\) to \(n^{-\frac{\beta}{K^*+2\beta}}\), where  \(\mathcal{D}=\prod_{j=1}^M P_j\) is the ambient dimension. This demonstrates that recovering the low-dimensional representation is crucial for mitigating the curse of dimensionality in tensor regression. Moreover, minimizing the derived generalization error bound yields explicit tuning-parameter choices for computationally lightweight tuning. Extensive simulations and real-data analyses show consistent performance across settings of varying complexity. Unlike black-box deep learning models, the framework supports direct visualization and interpretation of learned tensor representations without external post hoc tools.
	\end{abstract}

		\noindent%
\spacingset{1}		{\it Keywords: tensor features, weighted penalty, interpretability, prediction error, selection consistency, lightweight tuning.}
	\spacingset{1.8}

    	\section{Introduction}\label{sec_intro}
In the era of big data, multiway array data, such as  MRI scans, text, images, videos, traffic and social network such as user behavior data, have become increasingly common
and are typically represented as tensor form \citep{bhargava2015and, ma2019tensorized, panagakis2021tensor}.
However, tensor data analysis is challenging due to exponential dimensionality, complex multi-mode dependence, and the difficulty of achieving both flexibility and interpretability.

Early studies on tensor data analysis often treated each tensor element as a independent variable \citep{worsley2004unified, lazar2008statistical, penny2011statistical}. While this strategy is straightforward, it may overlook structural correlations among tensor elements and lead to loss of useful information. Another common approach is to vectorize tensors and then apply high-dimensional regression techniques \citep{cox2003functional, craddock2009disease}. However, vectorization destroys the multi-mode tensor structure and typically leads to ultra-high-dimensional regression problems, where estimation becomes difficult because of limited sample sizes and strong dependence among the resulting covariates.

Another line of work views tensor-valued covariates as discretized observations of underlying functions on continuous domains. For example, \citet{reiss2010functional, wang2014regularized} developed functional linear models for image-based prediction. These methods exploit spatial or temporal smoothness and therefore retain important structural information in image-like data. However, they usually require a meaningful continuous domain, together with prespecified notions of ordering, neighborhood, or smoothness. As a result, they may be less suitable for tensor covariates whose modes represent heterogeneous or discrete features, such as user behavior records, where natural adjacency may be unavailable.

To more effectively exploit the structural information in general tensor covariates, tensor linear regression models have been widely adopted, typically of the form $$f(\mX) = \langle\mX, \mB\rangle := \sum_{j_1,\cdots,j_M}X_{j_1,\cdots,j_M} B_{j_1,\cdots,j_M},$$
where $\mB \in \mathbb{R}^{P_1 \times \cdots \times P_M}$ is an unknown coefficient tensor and $B_{j_1,\ldots,j_M}$ denotes its $(j_1, \ldots, j_M)$-th element. To capture multi-mode dependencies, tensor decomposition techniques  such as CANDECOMP/PARAFAC (CP) \citep{harshman1970foundations} and Tucker decomposition \citep{tucker1966some} have been extensively  applied to $\mB$ \citep{guo2011tensor, zhou2013tensor, BTR2017, papadogeorgou2021soft, li2018tucker}. By imposing a low-rank structure on $\mB$, these approaches achieve substantial dimensionality reduction while preserving intrinsic dependencies across tensor modes.
To avoid pre-specifying the tensor rank, regularization-based methods have been developed \citep{zhou2014regularized, garvesh2019convex}, which adaptively learn low-dimensional representations. Despite these advances,  the method discussed above rely on linear modeling assumptions.

The linear assumption is often overly restrictive and frequently violated in real-world applications, leading to low predictive accuracy. Classical semiparametric and nonparametric models hence have been developed. For example, \cite{marx2011multidimensional} consider a single-index framework for tensor covariates, while \cite{hao2021sparse, chen2023bayesian, zhou2024broadcasted} study additive models applied to each element of the tensor covariate.
Product kernel methods  have also been proposed
\citep{signoretto2013learning, suzuki2016minimax, kanagawa2016gaussian, imaizumi2016doubly}
to  leverage the structural information of the tensor $\mX$. These approaches consider models ranging from  $f(\mX) = \sum_{v}\prod_{m=1}^M f_v^{(m)}(\vx_{m})$ with rank-1
to  more flexible model such as $f(\mX) = \sum_{v}\sum_{d=1}^D\prod_{m=1}^M f_v^{(m)}(\vx_{m,d})$ with rank-$D$, where $\mX = (\vx_1, \cdots, \vx_M)$ or $\mX= \sum_{d=1}^D\vx_{1,d}\circ\cdots\circ \vx_{M,d}$.
Although these regression models mitigate the curse of dimensionality in high-dimensional nonparametric settings, their predictive performance is often limited by strong model assumptions or inappropriate choices of  rank and kernel functions.

In recent years, deep neural networks (DNNs) have achieved remarkable success across various domains. Several studies \citep{chien2017tensor, kossaifi2020tensor, kolbeinsson2021tensor, DACOSTA2025130037} have incorporated tensor-based feature extraction into neural network architectures, leading to strong predictive performance for images, videos, and time-series data. However, these architectures are typically designed for tensor data with specific structures, such as spatial or temporal dependence, and may not directly apply to general tensor covariates with heterogeneous or unordered modes. Moreover, DNNs often require large training samples and usually provide limited  interpretability, which can restrict their use in scientific applications such as biology and medicine, where data are limited and interpretability is crucial.
To the best of our knowledge, there remains no existing tensor model that simultaneously achieves high predictive accuracy and  interpretability. This paper aims to develop a computationally tractable, highly predictive, and  interpretable framework that explicitly models the relationship between an $M$-way tensor covariate $\mX$ and a response variable $Y$.

We first consider a general tensor model:
\begin{eqnarray}
	\label{model}
	Y\sim g(\mX),
\end{eqnarray}
where $\mX \in \mathbb{R}^{P_1 \times \cdots \times P_M}$ is an $M$-way tensor predictor, $Y$ is a scalar response, and $\mathcal{D}:=\prod_{m=1}^M P_m$ denotes the ambient dimension of the tensor covariate. The function $g(\cdot)$ captures the relationship between the tensor covariate $\mX$ and the response $Y$. For example, when $Y$ is continuous, the model  reduces to $Y = g(\mX) + \varepsilon$ with measurement error $\varepsilon$, and when $Y$ is binary, the model  takes the form $P(Y = 1) = \text{Sigmoid}\{g(\mX)\}$.
To build a tensor model that is both highly predictive and intrinsically interpretable, we introduce a set of tensor feature extractors, denoted by $\mB_k$, $k=1,\ldots,K$, into the general model \eqref{model}. Specifically, we consider the following tensor neural network (TensorNN):
\begin{eqnarray}
	\label{model1}
	Y\sim
	g\big(\langle\mX,\mB_1\rangle,\cdots,\langle\mX,\mB_{K}\rangle\big),
\end{eqnarray}
where $g$  is approximated by DNNs. The  feature number $K$ can be chosen as large
as  the ambient dimension $\mathcal{D}$.
When \(K=\mathcal{D}:=\prod_{m=1}^M P_m\) and each \(\mB_k\) is chosen as a canonical indicator tensor that extracts one entry of \(\mX\), the vector \((\langle \mX,\mB_1\rangle,\ldots,\langle \mX,\mB_K\rangle)\) contains all entries of \(\mX\). In this case, model \eqref{model1} reduces to the general model \eqref{model}, showing that TensorNN retains the full predictive capacity of the general tensor regression framework.
Importantly, the introduction of $\mB_{k}, k\geq 1$  provides an opportunity to identify the pattern of how $\mX$  affects $Y$.
In particular, each $\mB_k$ serves as a feature extractor.
By identifying whether $\mB_k$ is zero or nonzero, and retaining only the corresponding nonzero features, we can determine the effective number of features in $\mX$ that contribute to $Y$.

In addition, excluding features associated with zero $\mB_k$ reduces the input dimensionality from $\mathcal{D}$ in model \eqref{model} to \(K\) in model \eqref{model1}. This reduction, in turn, leads to a more parsimonious network architecture for estimating $g$, which is automatically determined via the proposed architecture penalty. By adaptively reducing both the input dimensionality and the network size, the proposed method improves estimation stability and reduces stochastic error, while retaining  approximation capacity compared with the full model \eqref{model}. These effects, in turn, lead to reduced prediction error.
Moreover, the retained features $\langle\mX,\mB_ {k}\rangle$ provide the same interpretability as in a linear tensor regression model, offering explicit expressions of how the tensor covariate affects $Y$.

Our contributions can be  summarized as follows.
(1) To the best of our knowledge, we are the first to propose a  framework that simultaneously accounts for prediction accuracy, interpretability, and the structural information of tensor covariates.
(2) By imposing the group penalty on feature extractors $\mB_1, \mB_2, ..., \mB_K$, we  are able to identify \(K^*\), the minimal representation dimension  to ensure the approximation capacity needed for prediction.
(3) Theoretically, we establish the selection consistence for  \(K^*\) and  shows that the resulting TensorNN estimator achieves a non-asymptotic prediction error bound of order
\(C_* n^{-2\xi/(2\xi+K^*)}\), up to logarithmic factors, where \(\xi\) is the smoothness level of the low-dimensional function \(g\).
Thus, the convergence rate is governed by \(K^*\), rather than by the ambient tensor dimension \(\mathcal D\).
For  general $d$-dimensional function $f(\mX)$ estimated by deep neural networks, recent studies \citep{bauer2019deep, chen2019efficient, nakada2020adaptive, jiao2023deep} have established a convergence rate of $C_d n^{-\frac{2\xi}{2\xi + d^*}}$, where $d^*$ denotes a low-dimensional intrinsic dimension. However, the prefactor $C_d$ typically grows polynomially, or even exponentially, with the ambient  dimension $d$, which severely limits scalability in high-dimensional settings. In contrast, our convergence rate involves a prefactor $C_*$ that  depends on  $K^*$ and grows only logarithmically with the ambient tensor dimension $\mathcal{D}$.
This significantly mitigates the curse of dimensionality and enables our method to effectively handle ultra-high-dimensional nonparametric problems where $\mathcal{D} \gg n$.
(4) By minimizing the upper bound on the prediction error in Theorem~\ref{converge_thm1}, the relative penalty weights \(\vtau := (\tau_1,\tau_2,\tau_3)\), which control the penalties on feature indicators, feature components, and network size, can be explicitly specified. Consequently, after choosing a sufficiently rich working architecture, the main regularization tuning is reduced to a single parameter \(\lambda\), which controls the overall penalty strength. This leads to low computational complexity and a user-friendly practical implementation.
The effectiveness of this lightweight tuning strategy is demonstrated by TensorNN-op, which uses the theoretically motivated relative weights and achieves strong performance across the simulation and real-data studies in Tables~\ref{table_simu}, \ref{table_simu1}, and \ref{table_real}.
(5) Our model offers model-level interpretability through the learned tensor feature extractors, which act as synthesized tensor patterns over the covariate \(\mX\). In the simulation studies, the highlighted regions in Figure~\ref{heatmap_simu} closely align with the pre-defined active areas in Figure~\ref{active_region}, suggesting that the proposed method can recover meaningful underlying tensor patterns. In real-data applications, where predictive features are often irregular and spatially dispersed, the use of multiple tensor feature extractors allows TensorNN to capture complex structured signals, as illustrated in Figure~\ref{heatmap_real}. For example, in the smiling facial-image experiment, TensorNN highlights regions around the mouth, cheeks, and eyes as important predictive features. In contrast, for the mouth-closed classification task, TensorNN mainly highlights a localized and structured mouth region. These results provide interpretable tensor-level summaries of the learned representations.

The proposed method demonstrates strong predictive performance in both simulation studies and real data analyses. Across a wide range of experimental settings, including both simple and complex data-generating mechanisms, as well as diverse tensor types such as tabular and image data, our method consistently outperforms existing approaches. When applied to four real-world datasets, it delivers clear improvements over convolutional neural networks (CNNs) and traditional tensor regression methods, including STAR \citep{hao2021sparse} and TLR \citep{zhou2013tensor}, as shown in Tables~\ref{table_simu}, \ref{table_simu1}, and~\ref{table_real}.

The remainder of this paper is organized as follows. In Section \ref{sec_tnn_model}, we propose a penalized estimation method that simultaneously learns the feature extractors $\mB_k$,  their number $K$, and  the network  to approximate the function $g$ in the model (\ref{model1}). Section \ref{sec_theory}
establish the theoretical properties of the estimators, including the consistency of the features  dimension and
an upper bound of the generalization error. Sections \ref{sec:simu} and \ref{sec:real} present a series of simulation studies and four real-world applications.
All technical proofs are omitted for brevity.

\section{Tensor Neural Network and  Estimation}\label{sec_tnn_model}

Denote $[K]=\{1,\cdots, K\}$, $\mB_{[K]}=\{\mB_k: k\in[K]\}$, and $\langle\mX,\mB_{[K]}\rangle=\big(\langle\mX,\mB_1\rangle,\cdots,\langle\mX,\mB_K\rangle\big)$.
As mentioned in Section \ref{sec_intro}, a key strategy for building a highly predictive and  interpretable tensor regression model is to incorporate $\mB_k$s into the model (\ref{model}),  leading to the  framework in (\ref{model1}).
However, the tensors $\mB_{[K]}$ contain $K\mathcal{D}$ unknown parameters, typically far exceeding the sample size $n$. Such an ultrahigh-dimensional model imposes a significant computational burden and undermines the model's generalization capability \citep{Kawaguchi_Bengio_Kaelbling_2022}.
To reduce the number of parameters while capturing intrinsic dependencies among tensor covariates, we follow a standard approach in tensor regression and impose a CP decomposition \citep{zhou2013tensor, hao2021sparse, zhou2024broadcasted} on each \(\mB_k\) as follows:
\begin{align}\label{B_cp}
	\mB_{k} = \sum_{r = 1}^R \gamma_{kr}\vbeta_{kr1\bullet}\circ\cdots \circ\vbeta_{krM\bullet},
\end{align}
where $\vbeta_{krm\bullet}=(\beta_{krmj}, j\in[P_m])^T$ collects elements along the fourth (position) dimension,
`$\circ$'  denotes the vector outer product, and the rank $R$ is generally  much smaller  than $\min_j\{P_j\}$.
Under the decomposition \eqref{B_cp}, the number of parameters in $\mB_{[K]}$ is  reduced from $K\mathcal{D}$ to $K R \big(1 + \sum_{m=1}^M P_m\big)$, while preserving the intrinsic structural information of the tensor as much as possible \citep{kolda2009tensor}.
The parameters associated with the feature extractors $\mB_{[K]}$ are then defined as
\begin{align}\label{def_b}
	\vtheta_{fea} :=\{\vgamma_{1\bullet},\cdots,\vgamma_{K\bullet},\vbeta_{111\bullet},\cdots,\vbeta_{KRM\bullet}\big\}\in\bbR^{S_{fea}},
\end{align}
where $\vgamma_{k\bullet} = \{\gamma_{kj}, j\in [R]\}$, $S_{fea} = KR + KR\sum_{m = 1}^M P_m$ is the total number of the parameters
for the feature extractors.

We
approximate the target function $g$
by a feedforward neural network (FNN) with ReLU activation. A ReLU FNN with $L$  hidden layers is a mapping  $g : \mathbb{R}^{H_0} \to \mathbb{R}^{H_{L+1}}$ of the form $g(\vx) = T_L \circ \sigma \circ T_{L-1} \circ \cdots \circ \sigma \circ T_0(\vx)$,
where
$g_1 \circ g_2(\vx) := g_1(g_2(\vx))$,
the ReLU function
$\sigma(\vx):= \max(\vx, \vzero)$, applied elementwise, and
$T_l(\vx) := \mW_l\vx+\va_l$ is an affine transformation with $\mW_l \in \mathbb{R}^{H_{l+1} \times H_l}$ and $\va_l \in \mathbb{R}^{H_{l+1}}$ for $l=0, 1, ..., L$,   $H_l$  is the number of neurons in layer $l$.
We term $H=\max\{H_1, H_2, ..., H_L\}$ and $L$ the \textit{width} and \textit{depth} of the network. The \textit{size} $S_{net} = \sum_{l = 0}^LH_{l + 1}H_{l} + \sum_{l=0}^LH_{l + 1}$ is
the total number of parameters in the network.
Let $\mathcal{G}(K, H, L)$ be the class of ReLU networks with input dimension $K,$ width  $H$, and depth $L$.
Hence, the  parameters related to network are
\begin{align}\label{def_omega}
	\vtheta_{net} := \big\{\mW_0,\cdots,\mW_L,\va_0,\cdots,\va_L\big\}\in\bbR^{S_{net}}.
\end{align}

It is well established that the convergence rate of neural networks depends critically on the input dimension \citep{bauer2019deep}; in high-dimensional settings, this dependence may lead to the curse of dimensionality. In our framework, the representation dimension $K$ determines the input dimension of $g$ and thus controls the complexity of both the feature extractors $\mB_{[K]}$ and the subsequent network. If $K$ is over-specified, redundant components and network parameters inflate estimation variance and require larger sample sizes for accurate estimation. If $K$ is under-specified, the model lacks sufficient representation capacity, leading to increased approximation error. Therefore, selecting the minimal \(K\) that ensures sufficient representation capacity while minimizing stochastic error is crucial.

Moreover, theoretical analyses of neural networks \citep{schmidt2020nonparametric, tan2024generative} indicate that the size of the network $g$ should adapt to the feature dimension $K$ in order to achieve optimal convergence rates. A straightforward strategy is to search over a grid of candidate values of $K$ and network architectures, and then select the combination with the smallest prediction error. However, this strategy has two clear drawbacks. First, it is computationally expensive, since each candidate pair requires separate training and validation. Second, the selection is driven purely by empirical performance and does not directly provide theoretical guarantees.
To address these limitations, we propose a regularization approach that simultaneously
learns the feature dimension, the feature extractors, and  network  adapted to the learned dimension, while retaining theoretical guarantees.

Particularly, denote $\vtheta = \{\vtheta_{net}, \vtheta_{fea}\}$. Given i.i.d. samples $\mS^n\equiv {(\mX_{i},Y_{i})}_{i=1}^{n}$ of the population $\mS=(\mX, Y)$ with  a sample size  $n$,  we  estimate  $\vtheta$ by
\begin{align}\label{objective_f}
	\hat \vtheta \in \argmin_{\vtheta}\frac{1}{n}\sum_{i = 1}^n \ell\big(g_{{\small\vtheta_{net}}}\big(\langle\mX_i,\mB_{[K]}\rangle\big), Y_i\big) + \sP(\vtheta),
\end{align}
where
$g_{\vtheta_{net}}(\cdot)\in \sG(K, H, L)$ is  specified by $\vtheta_{net}$,
$\ell(\cdot,\cdot)$ is a  loss function  such as the squared loss for continuous responses and the logistic loss for binary responses,
\begin{align}\label{def_penalty2}
	\sP(\vtheta)
	=\lambda\Vert\vtheta\Vert_{\vtau}:=\lambda\big(\tau_1\sum_{k\in[K]}\Vert\vgamma_{k\bullet}\Vert_2 + \tau_2\sum_{(k,m,j)\in[K,M]}\Vert\vbeta_{k\bullet mj}\Vert_2 + \tau_3\Vert\vtheta_{net}\Vert_{1}\big),
\end{align}
$ \vbeta_{k\bullet mj} = (\beta_{krmj}, r\in [R])\trans$ collects elements along the rank dimension,
$%
[K,M] := \{(k, m, j)\mid k\in [K], j \in [P_m], m \in [M]\},$ and  $\|\cdot\|_v$ is the $v$-norm of a vector or a matrix.
The first penalty identifies the number of features  by detecting and excluding zero feature extractors in $\{\mB_k: k\in[K]\}$
because  $\vgamma_{k\bullet} = \vzero$ implies $\mB_k = \vzero$.
With the tensor expression in  (\ref{B_cp}), the second term imposes a group penalty on $\vbeta_{k\bullet mj}$,
the $j$-th slice along mode $m$ of $\mB_k$ across all CP components.
When $\vbeta_{k\bullet mj} =\vzero$, the corresponding $j$-th slice  of $\mX$ along mode $m$ is irrelevant to feature
$k$,  allowing us to identify the tensor covariates that are relevant to feature $k$. This, in turn, clarifies the  meaning of feature $k$, thereby enhancing interpretability and reducing model complexity.
The third term shrinks  redundant network weights towards $\vzero$,  controlling  the network
size to match the effective dimension,
i.e., the number of selected features.
Consequently, the automatic identifications of the number of features and network size
can be accomplished by appropriately tuning the parameters $\lambda$, $\tau_1, \tau_2$ and $\tau_3$.
In the next section, we derive explicit expressions for the relative weights $\vtau=(\tau_1,\tau_2,\tau_3)$ by minimizing an achievable upper bound on the prediction error established in Theorem~\ref{converge_thm1}. Consequently, only a single hyperparameter $\lambda$, which controls the overall regularization strength, needs to be tuned, substantially reducing the computational burden.

\section{Theoretical Analysis}\label{sec_theory}
Consider the true
function of $g$ in \eqref{model1} lies in the H\"older class $\sH^{\xi}\big([-D,D]^K, B_0\big)$
defined as
\begin{align*}
	\Big\{g:[-D,D]^{K}\rightarrow\bbR,\max_{\Vert\valpha\Vert_1\leq\lfloor\xi\rfloor}\Vert\partial^{\valpha}g\Vert_{\infty}\leq B_0,\max_{\Vert\valpha\Vert_1=\lfloor\xi\rfloor}\sup_{\vx\neq\vx^{'}}\frac{\vert\partial^{\valpha}g(\vx)-\partial^{\valpha}g(\vx^{'})\vert}{\Vert\vx -\vx^{'}\Vert_2^{\xi - \lfloor\xi\rfloor}}\leq B_0\Big\},
\end{align*}
with $\lfloor\xi\rfloor$ denoting the largest integer strictly smaller than $\xi$, where  $\partial^{\valpha} g=\partial^{|\valpha|} g/\partial x_1^{\alpha_1}\cdots\partial x_K^{\alpha_K}$ denotes the mixed partial derivative of order
$\valpha=(\alpha_1,\ldots,\alpha_K)^{T} \allowbreak \in \mathbb{N}_0^K$ with $|\valpha| = \sum_{i=1}^K\alpha_i$.

We estimate $g$ using a ReLU FNN. The corresponding function class is defined as
$
\sF(K,H,L)
:=
\left\{
g_{\vtheta_{\mathrm{net}}}\big(\langle \cdot,\mB_{[K]}\rangle\big):
g_{\vtheta_{\mathrm{net}}}\in \sG(K,H,L)
\right\},
$ which is also termed by TNN class in the paper.
For any function $f$, denote
$
\bbP\ell_f := \bbE\{\ell(f(\mX),Y)\}$ and
$\bbP\sL_f := \bbP(\ell_f-\ell_{f_0}),$
where $\ell_f:=\ell(f(\mX),Y)$  and $f_0$ denotes a true function as specified in Definition \ref{a1}. We further write
$
\|f\|_{L_2(\bbP_{\mX})}
:=
\{\bbE[f(\mX)^2]\}^{1/2},$
$
\|f\|_{\infty}:=\sup_{\mX}|f(\mX)|, $
$
\|\mC\|_{1,\infty}:=\max_j\sum_i |C_{ij}|,
$
and use $|\cdot|$ to denote the element-wise absolute value operation.
Since bias terms can be absorbed into the weight matrices by augmenting each layer input with a constant $1$, we assume without loss of generality that $\va_l=0$ for all layers $l$.

Throughout the paper, the subscript `$0$' denotes the true value, while the superscript `$*$' denotes the corresponding minimal quantity. Model \eqref{model1} is highly non-identifiable, due to  the flexibility in choosing $K$ and $\mB_{[K]}$ and the  non-identifiability of neural network parametrizations.
we therefore define the following class of true functions.

\begin{definition}[True function class] \label{a1}
	We define the true function class as
	$\mathcal{F}_0:=\{f_0: f_0 \in \argmin_{h\in \mathcal{F}}\bbE\big\{\ell\big(h\big(\mX\big), Y\big)\big\}\}$, where $\mathcal{F}=\underset{K\leq \mathcal{D}}{\cup}\big\{g\big(\langle\mX,\mB_{[K]}\rangle\big)\mid g\in \sH^{\xi}\big([-D,D]^K, B_0\big)\big\}$.
\end{definition}

Since the true function may admit  different numbers of extracted features, we introduce the following notion of minimal feature dimension.
\begin{definition}[Minimal feature dimension] \label{def:oid}
	The minimal feature dimension is defined as
	$K^*:= \min\big\{K:\ g(\langle \cdot,\mB_{[K]}\rangle) \in \sF_0 \big\}.$
\end{definition}

Then, the true oracle function class with known  $K^*$ is defined as
$$\sF_0^{or}:=\{g(\langle \cdot,\mB_{[K^*]}\rangle)\in\sF_0\},$$ and the corresponding oracle parameter space with known  $K^*$ is denoted as
\begin{align}\label{optimal_oracle}
	\vTheta^{or}(K^*, H, L):=\{\vtheta: \Vert f_{\vtheta} - f_0\Vert_{\infty}\leq \epsilon_n(f_0), f_{\vtheta}\in\sF(K^*, H, L), f_0\in\sF_0^{or}\},
\end{align}
where
$\epsilon_n(f_0) :=
\min_{f\in\sF(K^*, H, L)}
\Vert f - f_0\Vert_{\infty}.$
Then  $\epsilon_n:=\max\limits_{f_0\in \sF_0^{or}} \epsilon_n(f_0)$ is the approximation error between the true function class and the class of  network functions with the minimal number of features $K^*$.

\subsection{Nonasymptotic high-probability bounds}\label{sec_nonasym_bound}
To establish the  properties of the estimator, we introduce the following assumptions.
\begin{assumption}
	\label{bound_cond}(Boundedness).
	The covariate satisfies  $\mX\in [-1, 1]^{P_1\times\cdots\times P_M}$. There exist constants $C_x, B, F > 0$ such that $\Vert\text{vec}(\mX)\Vert_1\leq C_x,$ $\Vert\vtheta\Vert_{\infty}\leq B$, $\Vert f_0\Vert_{\infty}\leq F$, and $\Vert f\Vert_{\infty}\leq 2F$ for all $f\in \sF(K, H, L)$.
\end{assumption}
\begin{assumption}\label{lip_cond} (Lipschitz loss function).
	For any bounded functions $f_1$ and $f_2$, $\big\vert\ell\big(f_1(\bx),y\big) - \ell\big(f_2(\bx),y\big)\big\vert\leq \big\vert f_1(\bx) - f_2(\bx)\big\vert.$
\end{assumption}

\begin{assumption}\label{bern_cond}(Curvature). There exist constants $c_f,C_f>0$ such that for every $f\in \sF(K, H, L),$ $c_f\Vert f - f_0\Vert_{L_2(\bbP_{\mX})}^2\leq \bbP\sL_f \leq C_f\Vert f - f_0\Vert_{ L_2(\bbP_{\mX})}^2.$
\end{assumption}

Boundedness conditions in Assumption~\ref{bound_cond}  are standard in DNN theory
\citep{schmidt2020nonparametric, ohn2022nonconvex, meng2026inference}.
Assumption~\ref{lip_cond} requires the loss function to be Lipschitz, whereas
Assumption~\ref{bern_cond} imposes a Bernstein-type curvature condition that connects the excess risk with the prediction error.
Conditions analogous to Assumptions~\ref{lip_cond} and \ref{bern_cond} are widely used in empirical-risk analyses; see, for example,
\citet{bartlett2005local, koltchinskii2006local, alquier2019estimation}
for general nonparametric estimation theory and \citet{farrell2021deep} for neural network estimation.
The Lipschitz constant is normalized to $1$ without loss of generality, since rescaling the loss only changes the constants in the subsequent bounds.
Under these regularity conditions, we next identify a sparse subclass of models that achieves the desired approximation error.

\begin{lemma}\label{lemma1}
	Under Assumption \ref{bound_cond}, when $K^*$ is known, for any positive integer $l$ and
	$w_n=\left\lceil n^{K^*/\{2(2\xi+K^*)\}}\right\rceil$,
	the TNN class
	$\sF(K^*, H, L)$ with depth
	$
	L \geq
	L^*:= 21(\lfloor\xi\rfloor + 1)^2l\lceil \log_2(8l)\rceil + 2K^*
	$
	and width
	$
	H = H_1 = \cdots = H_L \geq
	H^*:= 38(\lfloor\xi\rfloor + 1)^23^{K^*}({K^*})^{\lfloor\xi\rfloor + 1}w_n\lceil \log_2(8w_n)\rceil
	$
	satisfies
	that $\epsilon_n\leq c_{\epsilon}n^{-\frac{\xi}{2\xi + K^*}}$, where $c_{\epsilon}=19 (2D)^{\xi}B_0(\lfloor\xi\rfloor + 1)^2({K^*})^{\lfloor\xi\rfloor + \max(\xi, 1)/2}l^{-2\xi/K^*}$.
\end{lemma}	
Lemma \ref{lemma1} implies that for any $f_0\in \sF_0^{or}$, there exists a TensorNN function  $f\in \sF(K^*, H, L)$  with network size $S_{net}$ satisfying    $S_{net}\geq S_{net}^*\asymp L^* (H^*)^2=O(n^{\frac{K^*}{2\xi + K^*}}\log_2^2n)$  such that $\Vert f - f_0\Vert_{\infty} = O(n^{-\frac{\xi}{2\xi + K^*}})$.

Denote
$J^* := \min\big\{\sum_{(k, m, j)\in[K, M]}1_{\{\vbeta_{k\bullet mj}\neq  \vzero\}}\ |\  g(\langle \cdot,\mB_{[K]}\rangle) \in \sF_0, K=K^*\big\}$ as the minimal number of feature components; $\sF^*$ as the sparse subclass of models with $K\leq K^*$,  $J\leq J^*$ and $S_{net}\leq S^*_{net}$;
and $\Vert\vtheta - \tilde\vtheta\Vert_{\vtau,\tilde\sI}:=
\tau_1\sum_{\tilde\gamma_{k\bullet}\neq \vzero}
\Vert\vgamma_{k\bullet} - \tilde\vgamma_{k\bullet}\Vert_2 +
\tau_2\sum_{\tilde\vbeta_{k\bullet mj}\neq\vzero}\Vert\vbeta_{k\bullet mj}- \tilde\vbeta_{k\bullet mj}\Vert_2
+ \tau_3\sum_{(\tilde\theta_{net})_j\neq 0}\vert(\theta_{net})_j -(\tilde\theta_{net})_j\vert$.

\begin{assumption}\label{RLL_cond}(Restricted lower Lipschitz condition).
	For any tuning vector $\vtau=(\tau_1,\tau_2,\tau_3)$ satisfying $\tau_i^{-1}=O(1)$, $i=1,2,3$, there exists a constant $\kappa_{\vtau}>0$ such that
	$
	\|f_{\vtheta}-f_{\tilde\vtheta}\|_{L_2(\bbP_{\mX})}
	\ge
	\kappa_{\vtau}
	\|\vtheta-\tilde\vtheta\|_{\vtau,\tilde\sI},
	$
	for any $f_{\vtheta}\in\sF(K,H,L)$ and $f_{\tilde\vtheta}\in\sF^*$ with representatives
	$
	\vtheta\in \operatorname{op}_{\vtau}(\vzero,f_{\vtheta}),
	\
	\tilde\vtheta\in \operatorname{op}_{\vtau}(\vtheta,f_{\tilde\vtheta}),
	$
	whenever the cone condition
	$
	\|\vtheta-\tilde\vtheta\|_{\vtau}
	\le
	\frac{2}{1-\eta}
	\|\vtheta-\tilde\vtheta\|_{\vtau,\tilde\sI},
	\  \eta\in(0,1),
	$
	is satisfied.
\end{assumption}
Here $\operatorname{op}_{\vtau}(\vtheta,f)$ denotes the set of parameterizations of the function $f$ that are closest to the reference parameter $\vtheta$ under the weighted norm $\|\cdot\|_{\vtau}$.
Both CP tensor decompositions and neural networks admit multiple parameterizations of the same function, due to scaling, permutation, and other invariances. Consequently, a global lower Lipschitz condition linking distances in the original parameter space to distances in the induced function space is generally impossible. Assumption~\ref{RLL_cond} can therefore be interpreted as a restricted local identifiability condition after accounting for the non-unique parameterizations of TensorNN functions. Specifically, it imposes the lower-Lipschitz-type requirement only after selecting a closest sparse representative through $\operatorname{op}_{\vtau}(\vtheta,f)$, and only along sparse cone directions.
In this sense, Assumption~\ref{RLL_cond} serves as a nonlinear analogue of the restricted eigenvalue (RE) condition in high-dimensional penalized regression \citep{10.1214/08-AOS620,bellec2018slope}. The RE condition is standard in penalized estimation theory: it rules out nontrivial sparse perturbations that produce little or no change in the fitted function. Similarly, Assumption~\ref{RLL_cond} ensures that, locally around the sparse oracle representation, the functional discrepancy is sufficiently informative about the relevant sparse parameter directions. This alignment between functional estimation and the parameter penalty enables the penalty to drive the estimator toward a sparse oracle representation.

Define
$M_n := \max_{l = 0,\cdots,L+1} \big(\big\Vert\vert\mW\vert_{L:l}\big\Vert_1\cdot \big\Vert\vert\mW\vert_{(l-2):-1}\big\Vert_{1,\infty}\big)$, where $|\mW|_{l:l'}:=|\mW_l|\cdots|\mW_{l'}|$, $\mW_{-1}=\mB$, and
$\mB=(\mathrm{vec}(\mB_1),\ldots,\mathrm{vec}(\mB_K))^\top$.
This quantity is used to control the complexity of $\sF(K,H,L)$.
Let $\pi_n$ denote a vanishing failure-probability bound.
The explicit expressions for $\pi_n$ and the constants $c_0$, $C_{\rho,\vtau}$, $\widetilde C_*$ and $\widetilde C_{\vtau}$ are omitted for brevity.

\begin{theorem}\label{converge_thm1}
	Suppose Assumptions~\ref{bound_cond}--\ref{RLL_cond} hold. Assume that
	$M$, $R$, and $\tau_i^{-1}$, $i=1,2,3$, are of constant order,
	$
	\max\{K,\max_{1\le m\le M}P_m,S_{\mathrm{net}}\}=O\{\operatorname{poly}(n)\},
	\
	L=O(L^*),
	$
	and the oracle parameters in $\vTheta^{or}(K^*,H^*,L^*)$
	satisfy $\max_l\{\Vert\mW_l^{or}\Vert_{1,\infty}, \Vert{\mW_l^{or}}\trans\Vert_{1,\infty}, \Vert\mB^{or}\Vert_{1,\infty}\}\leq F$.
	Further assume that
	$M_n=O(n),
	\
	J^*=O\left(n^{\frac{K^*}{2\xi+K^*}}\log_2^2 n\right).
	$
	If $n > \max\big\{R(K^* + J^*) + S^*_{net}, \frac{1}{r_{\vtau}^2}\max(1, \frac{201F^2\log n}{\sqrt{2}}, \frac{225F^2}{\sqrt{2}\log n})\big\}$ and the tuning parameter $\lambda$ satisfies
	$$
	\lambda\in
	\left(
	\frac{r_{\vtau}}{\eta}(2c_0+\delta_n),
	\,
	r_{\vtau}\{2(c_f-c_0)-\delta_n\}
	\right)$$  for $\eta\in (0,1)$,
	where $r_{\vtau}
	=
	\widetilde C_* n^{-\frac{\xi}{2\xi+K^*}}\log_2 n
	+
	\widetilde C_{\vtau} n^{-1/2}O\{\operatorname{poly}(\log n)\}$,
	$\delta_n=\{C_f\epsilon_n^2+\epsilon_n\sqrt{2\tilde\gamma/n}+4F\tilde\gamma/n\}/(C_{\rho,\vtau}r_{\vtau}^2)=o(1)$  with $\tilde\gamma=\widetilde C_*n^{\frac{K^*}{2\xi+K^*}}$,
	then
	\begin{equation}\label{thm1_converge}
		\|f_{\widehat\vtheta}-f_0\|_{L_2(\bbP_{\mX})}^2
		\le
		2C_{\rho,\vtau}r_{\vtau}^2
	\end{equation}
	with probability at least $1-\pi_n$.
\end{theorem}

Theorem~\ref{converge_thm1} provides an oracle-type prediction guarantee for TensorNN. Although the estimator is computed over the enlarged working class $\sF(K,H,L)$, its error bound adapts to the intrinsic sparse oracle structure. Specifically, the leading term in $r_{\vtau}^2$ is of order
$
n^{-2\xi/(2\xi+K^*)}
$
up to logarithmic factors, matching the classical minimax rate for $K^*$-dimensional $\xi$-smooth nonparametric regression \citep{stone1982optimal}. Thus, the effective nonparametric dimension is the learned feature dimension $K^*$, rather than the ambient tensor dimension $\mathcal{D}$.

Furthermore, the three-part penalty induces sparsity at different levels: the penalty on $\vgamma$ selects active tensor features, the penalty on $\vbeta$ identifies relevant tensor components within each feature, and the penalty on $\vtheta_{\rm net}$ regularizes the outer neural network. As a result, the stochastic error depends on the oracle sparsity level
$
s_n=R(K^*+J^*)+S^*_{\rm net},
$
rather than on the full number of parameters in the working model. Therefore, overspecified choices of $K$, $H$, and $S_{\rm net}$ affect the rate only through logarithmic factors, allowing the working model to be chosen sufficiently large for approximation while the penalty controls the effective model complexity.

A key characteristic of tensor regression is its ultra-high-dimensional nature, where the ambient tensor dimension
$\mathcal{D}$ greatly exceeds the sample size $n$.
Previous convergence theories for neural networks typically assume a  fixed or moderately growing input dimension. Recent work, such as \cite{bauer2019deep, chen2019efficient, nakada2020adaptive, jiao2023deep}, has derived rates of the form $C_dn^{-\frac{2\xi}{2\xi + d^*}}$ depending on an intrinsic dimension $d^*$.
However, the prefactor $C_d$ often exhibits polynomial or even exponential dependence on the nominal dimension $d$.
Such dependence can dominate the convergence rate when the nominal input dimension is the tensor ambient dimension $\mathcal{D} \gg n$. In contrast, Theorem \ref{converge_thm1} shows that,
the ambient tensor dimensions enter the stochastic error only through logarithmic factors such as $\log K$, $\log(\sum_{m=1}^M P_m)$, and $\log S_{\rm net}$, enabling effective application to ultra-high-dimensional tensor regression.

Theorem~\ref{converge_thm1} also provides practical guidance for tuning. The requirement \\
$
\lambda\in
\left(
\frac{r_{\vtau}}{\eta}(2c_0+\delta_n),
\,
r_{\vtau}\{2(c_f-c_0)-\delta_n\}
\right)$ in Theorem~\ref{converge_thm1}
shows that the global penalty level $\lambda$ should be chosen on the same order as $r_{\vtau}$, which reflects the intrinsic complexity of the TensorNN class. Thus, models with larger intrinsic complexity require stronger regularization to control stochastic error. This is consistent with the role of regularization parameters in other penalized estimators \citep{lecue2018regularization,alquier2019estimation}.
Moreover,
the explicit upper bound for $r_{\vtau}$ depends linearly on $1/\tau_1, 1/\tau_2, 1/\tau_3$. Hence, we can optimize the relative weights $\vtau=(\tau_1,\tau_2, \tau_3)$ to minimize $r_{\vtau}$,  resulting in:
\begin{corollary}\label{converge_coro}
	Under conditions in Theorem \ref{converge_thm1}, if  $S_{net} \asymp \operatorname{poly}(K)$ and $P_m \asymp \operatorname{poly}(K)$ uniformly over $m$,
	the rate term $r_{\vtau}$ is minimized, up to a logarithmic factor $\log\nu$, by
	the relative weights $\vtau = (\tau_1, \tau_2, \tau_3)$, where
	\begin{align}\label{relative_weights}
		\tau_1 = \frac{\sqrt{\nu_1}}{\sqrt{\nu_1} + \sqrt{\nu_2} + \sqrt{\nu_3}}, \tau_2 = \frac{\sqrt{\nu_2}}{\sqrt{\nu_1} + \sqrt{\nu_2} + \sqrt{\nu_3}}, \tau_3 = \frac{\sqrt{\nu_3}}{\sqrt{\nu_1} + \sqrt{\nu_2} + \sqrt{\nu_3}},
	\end{align}
	with $\nu_1 = \sqrt{R}\log\big\{2RK\cdot\nu\big\}, \nu_2 = \sqrt{R}\log\big\{2RK(\sum_{m=1}^MP_m)\cdot\nu\big\}$, and $\nu_3 = \log\big\{2S_{net}\cdot\nu\big\}$, for some $\nu \geq 1 + C_{\nu}nM_n$ with $C_{\nu}$ depending on $L$.
\end{corollary}

\begin{corollary}\label{converge_coro11}
	Under the conditions of Corollary \(\ref{converge_coro}\), assume further that \(\nu\)  in   \(\eqref{relative_weights}\) is of the order $O(\text{poly}(n))$. Then the \(K^*\)-dependent optimal rate \(\eqref{thm1_converge}\) holds for any
	positive weights \((\tau_1,\tau_2,\tau_3)\) satisfying \(\tau_1+\tau_2+\tau_3=1\), \(\tau_i^{-1}=O(1)\), \(i=1,2,3\), or
	$\tau_1 \geq  \frac{\sqrt{\nu_1}}{\sqrt{\nu_1}+\sqrt{\nu_2}+\sqrt{\nu_3}},$
	with \(\nu_1,\nu_2,\nu_3\) defined in \(\eqref{relative_weights}\).
\end{corollary}

Theorem~\ref{converge_thm1} and Corollaries~\ref{converge_coro}
and~\ref{converge_coro11} motivate the lightweight tuning strategy developed in the paper: instead of jointly searching over
multiple penalty parameters and network architectures, TensorNN can be
implemented with theory-guided relative penalty weights and  a sufficiently rich working architecture, leaving only  a rough selection of network depth $L = O(L^*)$, and the overall regularization strength $\lambda$ to  be tuned.
The effectiveness of this strategy is supported by the empirical results in Sections~\ref{sec:simu} and~\ref{sec:real}. In particular,
TensorNN-op, which uses the relative penalty weights prescribed by
Corollary~\ref{converge_coro}, achieves consistently strong performance
in both simulation studies and real data analyses.

	\subsection{The consistency of feature selection}\label{sec_consist}

	To establish the selection consistency of $K^*$, we further introduce Assumptions  \ref{bound_cond2} and \ref{identi_cond}.

	\begin{assumption}(Derivative regularity)\label{bound_cond2}
		\begin{itemize}
			\item  For any $|\theta|\le 2F$, the loss function satisfies
			$
			|\partial_1^2\ell(\theta,y)|\le C_\ell .
			$
			Moreover, conditional on $\{\mX_i\}_{i=1}^n$, the score variables
			$\partial_1\ell(f_0(\mX_i),Y_i), i=1,\ldots,n,$
			are independent, mean-zero, and sub-Gaussian with variance proxy $\sigma_e^2$,
			where $\partial_1 \ell(\cdot,\cdot)$ and $\partial_1^2 \ell(\cdot,\cdot)$ denote the first- and second-order partial derivatives of the loss function $\ell$ with respect to its first argument, respectively.

			\item   Let $\vtheta^*$ be an optimal sparse TensorNN parameter such that
			$f_{\vtheta^*}\in\mathcal F^*$ and $f_{\vtheta^*}=f_{\vtheta^{\rm or}}$, where
			$\vtheta^{\rm or}$ is defined in  \eqref{optimal_oracle}. For any $\vtheta$ satisfying
			$\|\vtheta-\vtheta^*\|_{\vtau}=O(r_{\vtau})$, the Clarke subdifferentials of
			$g_{\vtheta_{\rm net}}$ with respect to the $k$-th feature input satisfy
			$
			\left\{
			\frac1n\sum_{i=1}^n
			\big|
			\partial_k g_{\vtheta_{\rm net}}(\langle \mX_i,\mB_{[K]}\rangle)
			\big|^2
			\right\}^{1/2}
			\leq C_g,  k=1,\ldots,K.
			$
			In addition, for all perturbations $\Delta_1,\ldots,\Delta_K$ satisfying
			$\sum_{k=1}^K|\Delta_k|=O(r_{\vtau})$, the local variation condition
			\[
			\sum_{k=1}^K
			\left\{
			\frac1n\sum_{i=1}^n
			\left|
			\partial_k g_{\vtheta_{\rm net}}
			(\langle \mX_i, \mB_{[K]}\rangle+\ve_k\Delta_k)
			-
			\partial_k g_{\vtheta_{\rm net}}
			(\langle \mX_i, \mB_{[K]}\rangle)
			\right|^2
			\right\}^{1/2}
			\le C_g
			\]
			holds, where $\ve_k$ denotes the $k$-th canonical basis vector.
		\end{itemize}
	\end{assumption}

	Assumption \ref{bound_cond2} consists of two regularity requirements. The first one imposes local smoothness
	of the loss and a sub-Gaussian score condition at the truth. These are standard assumptions in high-dimensional generalized  nonparametric regression models \citep{haris2022generalized,yang2022ultrahigh}. For example, for generalized nonparametric models with
	$
	\ell(\theta,y)=-\{y\theta-\psi(\theta)\},\ \psi'(f_0(\mX_i))=E(Y_i\mid \mX_i),
	$
	the score $\partial_1\ell(f_0(\mX_i),Y_i)$ has conditional mean zero, and the curvature term $\partial_1^2\ell(\theta,y)=\psi''(\theta)$ is bounded on compact domains for common models.
	The second part of Assumption \ref{bound_cond2} controls the sensitivity of the outer neural network with respect to each tensor feature. Since ReLU networks are locally Lipschitz but not differentiable everywhere, the Clarke subdifferential provides the appropriate generalized derivative \citep{clarke1990optimization}. The boundedness and local variation conditions on these subdifferentials ensure that small perturbations of the tensor features do not produce unstable changes in the feature-level score. These are local conditions around the sparse oracle representation, and they become less restrictive as $r_{\vtau}\to 0$.

	\begin{assumption}(Feature signal strength)\label{identi_cond}
		There exists a sufficiently large constant $C_0>0$ such that, for any $\widetilde K<K^*$, any tensor feature parameters $\mB_{[\widetilde K]}$, and any measurable function $g:\mathbb R^{\widetilde K}\to\mathbb R$,
		$
		\left\{
		\frac1n\sum_{i=1}^n
		\left[
		f_0(\mX_i)-g\big(\langle \mX_i, \mB_{[\widetilde K]}\rangle\big)
		\right]^2
		\right\}^{1/2}
		\ge C_0r_{\vtau}.
		$
	\end{assumption}

	Assumption \ref{identi_cond}  requires that the true regression function cannot be approximated, even empirically, by any model using fewer than $K^*$ tensor features. Thus, each of the $K^*$ oracle features contributes non-redundant predictive information. This condition is analogous to the beta-min or signal separation assumptions commonly imposed for model selection consistency in sparse regression \citep{zhao2006model, huang2010variable}. Without such a condition, feature selection consistency is not identifiable: if $f_0$ can be represented equally well with fewer than $K^*$ features, then no method can consistently distinguish $K^*$ from a smaller number of effective features.

	\begin{theorem}\label{selection_thm2}
		Suppose that the conditions of Corollary \(\ref{converge_coro}\), together with Assumptions \(\ref{bound_cond2}\) and \(\ref{identi_cond}\), hold. Then there exists a constant
		$C=C(\eta,B,M,R,C_{\ell},\sigma_e)>0$  such that, if \(\tau_1 \geq C C_g^2\), then
		$ \bbP(\widehat K = K^*) \to 1$
		as \(n\to\infty\).
	\end{theorem}

	Since the selection of nonzero components in \(\vbeta_{k\bullet mj_m}\) and \(\theta_j\) is conditional on the selection of \(K\),   selection consistency requires the feature-level penalty to dominate the penalties on tensor components and outer neural network parameters. That is, \(\tau_1 \geq C C_g^2\) or equivalently, $\tau_1/\tau_2$ and $\tau_1/\tau_3$ should be  large. Moreover, if they  remain of constant order, as required in
	Theorems~\ref{converge_thm1}, Theorem \ref{selection_thm2} establishes that TensorNN enjoys both oracle-rate prediction performance and selection consistency for $K^*$.
	Theorem \ref{selection_thm2} also reveals an important practical advantage of TensorNN. In standard sparse parametric models, increasing the penalty strength to improve selection consistency often introduces additional estimation bias, leading to a tradeoff between prediction accuracy and variable selection \citep{zhao2006model, Sara2008high}. In contrast, TensorNN benefits from its non-unique parameterization: the scale of the feature indicator \(\vgamma_{k\bullet}\) can be offset by the tensor components and the parameters of the outer neural network. As a result, \(\tau_1\) can be increased to encourage feature selection without compromising the oracle prediction rate, as long as the relative penalty weights remain of constant order and the required boundedness conditions are satisfied.  Therefore, Theorem \ref{selection_thm2} provides a theoretical justification for the feature-selection version of TensorNN, showing that it can simultaneously recover the true number of active features \(K^*\) and attain the optimal convergence rate.

	\section{Simulation Studies}\label{sec:simu}

	In this section, we evaluate the proposed TensorNN on several tensor regression and classification tasks. We consider two variants of the proposed method. TensorNN-op uses the theoretically motivated relative penalty weights suggested by Corollary \ref{converge_coro}, with emphasis on
	prediction performance. TensorNN-sc further tunes the feature-selection weight \(\tau_1\), motivated by Corollary \ref{converge_coro11} and Theorem \ref{selection_thm2}, to ensure  selection consistency. When no distinction is needed, we refer to both variants simply as TensorNN. We compare TensorNN with three competing methods: the convolutional neural network \citep[CNN,][]{li2021survey}, nonlinear sparse tensor additive regression with a group lasso penalty \citep[STAR,][]{hao2021sparse}, and tensor linear regression with a lasso penalty \citep[TLR,][]{zhou2013tensor}.

	\subsection{Simulation Settings}\label{sec_simu_set}

	\textbf{Scenario 1.} We generate multi-way tensor covariates
	\(\mX\in\mathbb{R}^{P_1\times\cdots\times P_M}\), with \(M=2\) or \(3\).
	The true regression function is defined as
	$
	f_0(\mX)=\{f_{\rm raw}(\mX)-\mu_0\}/\sigma_0,
	$
	where
	$
	f_{\rm raw}(\mX)
	=h_0\{h_1(\mX_{\sA_1}),\ldots,h_{K^*}(\mX_{\sA_{K^*}})\},
	$
	$
	h_k(\mX_{\sA_k})
	=
	\langle \mB_k, \mX\rangle,
	$
	the coefficient tensor \(\mB_k\) is supported on \(\sA_k\) and $\Vert\mB_k\Vert_F = 1$.
	Thus, \(h_k\) extracts local information from the active  subregion \(\sA_k\)
	and \(h_0\) combines these local features.
	For least-squares regression, we set
	\(\mu_0=0\) and \(\sigma_0=1\). For logistic regression, where
	\(\Pr(Y=1\mid \mX)=\mathrm{sigmoid}\{f_0(\mX)\}\), the constants \(\mu_0\) and
	\(\sigma_0\) are chosen so that the resulting conditional probabilities cover
	a broad range of \([0,1]\).

\begin{figure}[t]
		\centering
		\includegraphics[width = \textwidth]{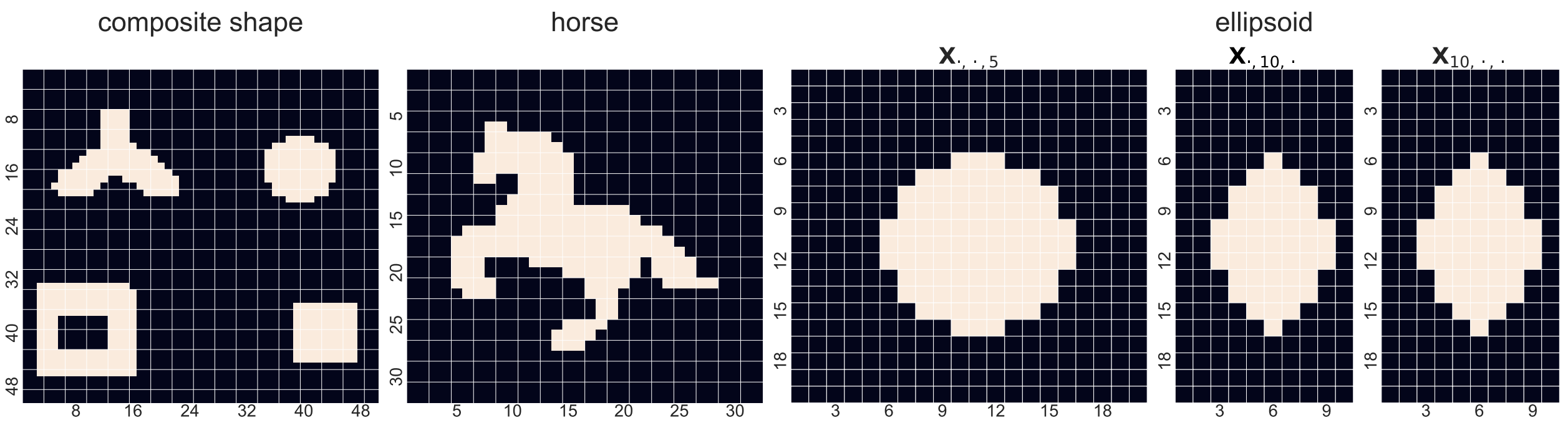}
		\spacingset{1}
		\caption{
			Active regions for the three groups of simulation settings. The first and second heatmaps correspond to the composite and horse-shaped active regions, respectively. The three rightmost heatmaps show three slices of the ellipsoidal active region, corresponding to \(\mX_{\bullet,\bullet,5}\), \(\mX_{\bullet,10,\bullet}\), and \(\mX_{10,\bullet,\bullet}\), respectively. Bright cells indicate active tensor entries.
		}
		\label{active_region}
	\end{figure}

	The six simulation settings comprise three active-region geometries: a composite pattern on \(50\times50\) tensors, a horse-shaped region on \(32\times32\) tensors, and an ellipsoidal region on \(20\times20\times10\) tensors, as shown in Figure~\ref{active_region}. For each geometry, we consider an additive setting and an interaction setting, in which \(h_0\) combines the extracted local features additively or through interactions, respectively. The simulation specifications are summarized in Table~\ref{table_setting}.

	\begin{table}[t]
		\centering
		\spacingset{1}
		\caption{
			The specifications of simulation settings. The six settings are organized into three groups by their shapes of the active regions.}\label{table_setting}
		\spacingset{1.3}
		\resizebox{0.9\textwidth}{!}{	\begin{tabular}
				{ |c|  C{1.8cm}| C{1.8cm}|   C{1.8cm} |C{1.8cm}|   C{1.8cm} |C{1.8cm} | }
				\hline
				Setting ID & 1 & 2  & 3 & 4 & 5   & 6 \\
				\hline
				Active region pattern & \multicolumn{2}{c|}{Composite shape} & \multicolumn{2}{c|}{Horse} & \multicolumn{2}{c|}{Ellipsoid}\\
				\hline
				Regression type & \multicolumn{4}{c|}{least squares
				} & \multicolumn{2}{c|}{logistic
				}\\
				\hline
				Model type& additive & interaction & additive & interaction &additive & interaction \\
				\hline
				True function $f_0$& \multicolumn{6}{c|}{$f_0(\mX) = \big\{h_0\big(h_1(\mX_{\sA_1}), \cdots, h_{K^*}(\mX_{\sA_{K^*}})\big) -\mu_0\big\}/\sigma_0$}\\
				\hline
				$h_k$ & \multicolumn{6}{c|}{$h_k(\mX_{\sA_k}) = \big\langle \mB_k, \mX\big\rangle$ }\\
				\cline{2-7}
				$\mu_0$ & \multicolumn{4}{c|}{ $0$}& \multicolumn{2}{c|}{calibrated}\\
				\cline{2-7}
				$\sigma_0$ & \multicolumn{4}{c|}{ $1$}& \multicolumn{2}{c|}{calibrated}\\
				\cline{2-7}		
				$K^*$ &  $5$&   $5$  & $6$&   $6$   &  $4$&  $4$\\
				\hline
			\end{tabular}
		}
	\end{table}

	\textbf{Scenario 2} is similar to Scenario 1, except that the  feature extractor is nonlinear, that is,
	$
	h_k(\mX_{\sA_k}) = \big\langle \mB_k, \sin(\pi\mX/2)\big\rangle,
	$
	where the sine transformation is applied elementwise.
	This scenario allows us to examine a broader model class than the exact TensorNN representation in \eqref{model1}.

	For each setting, the entries of \(\mX\) are independently generated from \(\mathrm{Unif}(-1,1)\). In least squares settings, \(Y=f_0(\mX)+\epsilon\), where \(\epsilon\sim N(0,\sigma_\epsilon^2)\) and \(\sigma_\epsilon^2\) is chosen so that \(\mathrm{Var}\{f_0(\mX)\}/\mathrm{Var}(\epsilon)=50\). In logistic settings, \(Y\sim \mathrm{Bernoulli}[\mathrm{sigmoid}\{f_0(\mX)\}]\). We conduct 50 independent replications, each with 2,000 training, 400 validation, and 1,600 test observations.
	The validation set is used for selecting tuning parameters $\lambda$ and $\tau_1$.  The initial feature dimension \(K\), CP rank \(R\), network architecture  and relative weights $\vtau$  are selected following the theory-guided or  trial-based procedures.

	We evaluate prediction by relative prediction error (RPE) for least squares regression and misclassification rate (MCR) for classification:
	$
	\mathrm{RPE}
	=
	\frac{\sum_{i\in S}\{y_i-\widehat f(\mX_i)\}^2}
	{\sum_{i\in S}y_i^2}$ and $
	\mathrm{MCR}
	=
	\frac{1}{|S|}\sum_{i\in S}|y_i-\widehat y_i|,
	$
	where \(S\) denotes the test set and
	$
	\widehat y_i=I\{\mathrm{sigmoid}(\widehat f(\mX_i))\geq 0.5\}.
	$
	We also report model estimation error (MEE). For least squares regression, MEE is the relative \(L_2(\bbP_{\mX})\) error $
	\|\widehat f-f_0\|_{L_2(\bbP_{\mX})}^2/\|f_0\|_{L_2(\bbP_{\mX})}^2.
	$
	For classification, MEE is defined as
	$
	\mathbb{E}\big[\{\widehat{\mathbb{E}}(Y\mid \mX)-\mathbb{E}(Y\mid \mX)\}^2\big],
	$
	where \(\widehat{\mathbb{E}}(Y\mid \mX)=\mathrm{sigmoid}\{\widehat f(\mX)\}\).

	The optimization and tuning procedures are kept comparable across all numerical experiments and real-data analyses. For TensorNN, we solve the penalized empirical-risk problem in \eqref{objective_f} using the RMSProp algorithm \citep{hinton2012neural}, which jointly updates the tensor feature extractors \(\{\mB_k\}_{k=1}^K\) and the outer neural network \(g_{\vtheta_{net}}\). The competing methods are trained under the same optimization scheme whenever applicable.
	The batch size is set to $0.2 N_{training}$ where $N_{training}$ is the sample size of training set. The default RMSProp hyperparameters are used, with \(\alpha=0.99\), \(\epsilon=10^{-8}\), zero weight decay, zero momentum, and \(\mathrm{centered}\) is set to False.
	The learning rate is selected from \(10^{-5}\) to \(10^{-3}\) according to convergence behavior.
	For each candidate tuning-parameter setting, training is monitored at 10 equally spaced checkpoints, and the checkpoint with the smallest validation error is selected.
	All computations were implemented in Python.

	\subsection{Simulation Results}\label{simu_result}

	\textbf{Scenario 1.}
	Table~\ref{table_simu} reports the average RPE/MCR and MEE with SD, and  the median selected feature number \(\widehat K\) with MAD, over 50 replications.
	TensorNN achieves the best performance in all settings except Setting 1, where TLR is slightly better. This is likely because the true function \(h_0(\vx)=\sum_{k=1}^5 x_k+\ln(x_k^2+2)/k\)
	is dominated by its linear component and is therefore closer to the TLR structure.
	Comparing the interaction settings (Settings 2, 4, and 6) with their additive counterparts (Settings 1, 3, and 5),
	we observe that the advantage of TensorNN becomes more pronounced as the true function becomes more complex.
	Moreover, TensorNN-sc generally recovers \(K^*\) well, whereas TensorNN-op tends to slightly overestimate it, especially in Settings 5 and 6, where the active regions associated with different local features begin to overlap.

	\begin{table}[t]
		\centering
		\spacingset{1}
		\caption{
			Average RPE/MCR and MEE, together with the median selected feature number \(\widehat K\), for Scenario 1. Numbers in parentheses are standard deviations for RPE/MCR and MEE, and MADs for \(\widehat K\), over 50 replications. Boldface indicates the best prediction or estimation performance in each setting.
		}
		\spacingset{1.3}
		\resizebox{\textwidth}{!}{
			\begin{tabular}{ c| c c| c c | c c}
				\hline
				Setting ID & 1 & 2 & 3 & 4 & 5 & 6\\
				\hline
				Active region type & \multicolumn{2}{c|}{Composite shape} & \multicolumn{2}{c|}{Horse} & \multicolumn{2}{c}{Ellipsoid}\\
				\hline
				Regression type & \multicolumn{4}{c|}{least squares } & \multicolumn{2}{c}{logistic }\\
				\hline
				Model type& additive & interaction & additive & interaction &additive & interaction \\
				\hline
				\hline
				& \multicolumn{4}{c|}{$\text{RPE}$} & \multicolumn{2}{c}{$\text{MCR}$} \\
				\cline{2-7}
				TensorNN-op & $0.0483 (0.0005)$ & $0.0690 (0.0017)$ & $\mathbf{0.0420} (0.0003)$ & $\mathbf{0.0536} (0.0006)$ & $0.1705 (0.0013)$ & $\textbf{0.2069} (0.0017)$ \\
				TensorNN-sc & $0.0476 (0.0004)$ & $\mathbf{0.0647} (0.0012)$ & $\mathbf{0.0421} (0.0003)$ & $\mathbf{0.0537} (0.0005)$ & $\mathbf{0.1680} (0.0015)$ & $\mathbf{0.2069} (0.0018)$ \\
				CNN & $0.1156 (0.0019)$ & $0.1348 (0.0021)$ & $0.1122 (0.0007)$ & $0.1349 (0.0009)$ & $0.1843 (0.0019)$ & $0.2276 (0.0022)$ \\
				STAR & $0.0549 (0.0004)$ & $0.2054 (0.0017)$ & $0.0743 (0.0005)$ & $0.1291 (0.0009)$ & $0.1748 (0.0022)$ & $0.2223 (0.0019)$ \\
				TLR & $\mathbf{0.0457} (0.0003)$ & $0.1524 (0.0010)$ & $0.0648 (0.0005)$ & $0.1061 (0.0007)$ & $0.1781 (0.0015)$ & $0.2257 (0.0019)$ \\
				\hline
				\hline
				& \multicolumn{6}{c}{$\text{MEE}$} \\
				\cline{2-7}
				TensorNN-op & $0.0128 (0.0001)$ & $0.0476 (0.0015)$ & $\mathbf{0.0227} (0.0002)$ & $\mathbf{0.0157} (0.0002)$ & $0.0631 (0.0016)$ & $\mathbf{0.0948} (0.0022)$ \\
				TensorNN-sc & $0.0125 (0.0001)$ & $\mathbf{0.0435} (0.0010)$ & $\mathbf{0.0227} (0.0002)$ & $\mathbf{0.0158} (0.0002)$ & $\mathbf{0.0591} (0.0019)$ & $\mathbf{0.0944} (0.0020)$ \\
				CNN & $0.0425 (0.0007)$ & $0.1096 (0.0017)$ & $0.0937 (0.0004)$ & $0.0540 (0.0004)$ & $0.0981 (0.0024)$ & $0.1086 (0.0017)$ \\
				STAR & $0.0157 (0.0001)$ & $0.1773 (0.0009)$ & $0.0554 (0.0003)$ & $0.0515 (0.0002)$ & $\mathbf{0.0588} (0.0022)$ & $0.1131 (0.0008)$ \\
				TLR & $\mathbf{0.0116} (0.0001)$ & $0.1260 (0.0005)$ & $0.0458 (0.0003)$ & $0.0410 (0.0002)$ & $\mathbf{0.0585} (0.0007)$ & $0.1145 (0.0004)$ \\
				\hline
				\hline
				& \multicolumn{6}{c}{Median of $\widehat K$ (MAD)} \\
				\cline{2-7}
				TensorNN-op & $5.0 (1.0)$ & $5.0 (1.0)$ & $7.0 (2.0)$ & $6.0 (1.0)$ & $12.5 (8.5)$ & $8.0 (4.0)$ \\
				TensorNN-sc & $4.0 (1.0)$ & $4.0 (1.0)$ & $7.0 (2.0)$ & $6.0 (1.0)$ & $6.0 (4.0)$ & $5.0 (3.5)$ \\
				\hline
		\end{tabular} }
		\label{table_simu}
	\end{table}

	The comparison with CNN further illustrates the distinction between local convolutional feature learning and adaptive tensor feature extraction. As shown in Table~\ref{table_simu}, TensorNN achieves the best overall performance across the simulation settings, whereas CNN often gives the worst or near-worst results.

	This performance gap may be attributed to the fact that these settings require aggregating information from spatially extended or dispersed active regions, whereas the CNN considered here relies primarily on local receptive fields. As a result, such a CNN may be less effective at combining spatially separated tensor signals, especially when the sample size is limited. In contrast, TensorNN directly learns sparse tensor feature extractors through the structured penalty,
	allowing it to adapt to both dispersed additive structures and spatially extended active regions, as well as nonlinear interaction patterns.

	The similar RPE/MCR and MEE values of TensorNN-op and TensorNN-sc in Table~\ref{table_simu} suggest
	that the theory-guided lightweight tuning scheme of TensorNN-op already achieves strong generalization performance. The main difference lies in feature selection. By further tuning \(\tau_1\), TensorNN-sc imposes stronger feature-level sparsity and selects a more parsimonious representation, with \(\widehat K\) closer to the oracle feature dimension \(K^*\). By pruning redundant feature extractors, TensorNN-sc slightly improves the prediction and estimation accuracy of TensorNN-op. Empirically, TensorNN-sc selects substantially larger \(\tau_1\) values than TensorNN-op, which is consistent with Theorem~\ref{selection_thm2}: appropriately increasing the feature-level penalty promotes selection consistency of the tensor features.

\begin{figure}[t]
		\centering
		\includegraphics[width = \textwidth]{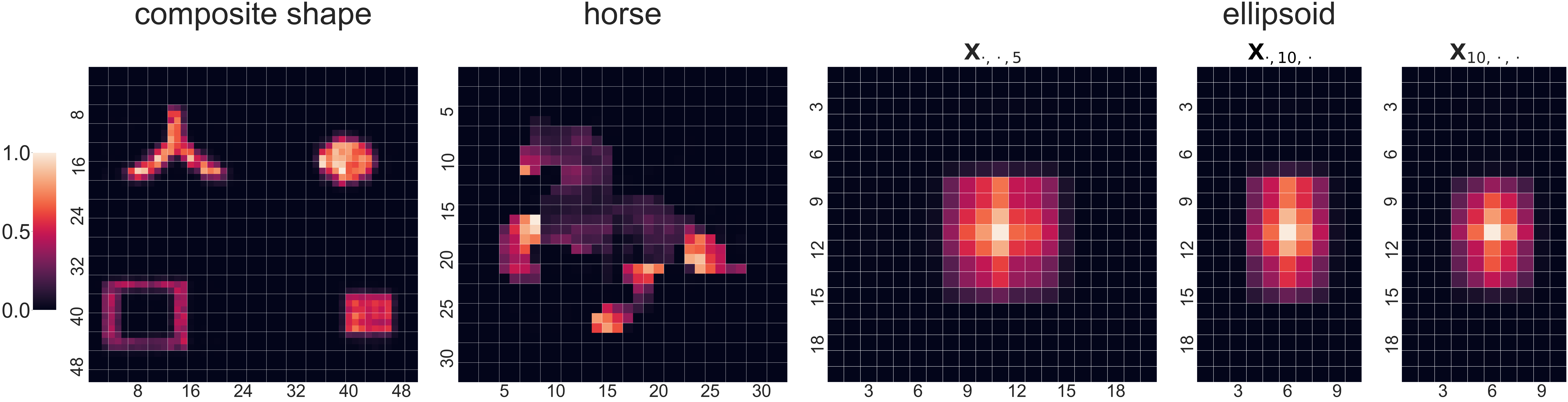}
		\spacingset{1}
		\caption{The heatmaps obtained by TensorNN-sc under the interaction settings (Settings 2, 4, and 6). Columns 1 and 2 correspond to the composite pattern (Setting 2) and the horse-shaped pattern (Setting 4), respectively. Columns 3--5 exhibit three slices of the estimated active regions corresponding to $\mX_{\bullet,\bullet, 5}$, $\mX_{\bullet,10, \bullet}$, and $\mX_{10,\bullet, \bullet}$ under Setting 6, respectively. }
		\label{heatmap_simu}
	\end{figure}

	Finally, we visualize the estimated active regions to assess the interpretability of TensorNN. We combine the estimated feature extractors \(\widehat{\mB}_k\), weighted by the outer neural network parameters, and then sum the element-wise squares of these weighted tensors to form the interpretable tensor \(\widehat{\mathbf V}^2\). Figure~\ref{heatmap_simu} displays the estimated active regions under the interaction settings, with each heatmap selected from the replicate with the smallest RPE.
	TensorNN-sc accurately recovers the composite active pattern in Setting 2.
	In the more complex horse setting (Setting 4), the highlighted regions are largely consistent with the true active regions shown in Figure~\ref{active_region}.
	For the ellipsoid setting (Setting 6), TensorNN-sc captures the intended pattern with strong central activation gradually weakening toward the boundary. These visualizations demonstrate that TensorNN-sc can recover meaningful active tensor regions while maintaining strong predictive performance. The heatmaps for Settings 1, 3, and 5 (omitted for brevity) lead to similar conclusions.

	We also conduct robustness experiments on network width; the results (omitted for brevity) show that network width has only a marginal effect on model estimation, consistent with the logarithmic dependence suggested by Theorem \ref{converge_thm1}.

	\textbf{Scenario 2.}
	Table~\ref{table_simu1} reports the results for Scenario 2, where the local feature extractors are nonlinear in the tensor covariates. The overall conclusions are consistent with those from Scenario 1: TensorNN maintains strong prediction and estimation performance across the six settings. These results further demonstrate the generality and adaptivity of TensorNN, suggesting that the proposed framework can accommodate a broad range of tensor structures and underlying regression functions beyond the exact model representation in \eqref{model1}.

	\begin{table}[t]
		\centering
		\spacingset{1}
		\caption{Average RPE/MCR and MEE for Scenario 2. Numbers in parentheses are standard deviations over 50 replications. Boldface indicates the best prediction or estimation performance in each setting.
		}
		\spacingset{1.3}
		\resizebox{\textwidth}{!}{
			\begin{tabular}{ c| c c| c c | c c}
				\hline
				Setting ID & 1 & 2 & 3 & 4 & 5 & 6\\
				\hline
				Active region type & \multicolumn{2}{c|}{Composite shape} & \multicolumn{2}{c|}{Horse} & \multicolumn{2}{c}{Ellipsoid}\\
				\hline
				Regression type & \multicolumn{4}{c|}{least squares} & \multicolumn{2}{c}{logistic}\\
				\hline
				Model type& additive & interaction & additive & interaction &additive & interaction \\
				\hline
				\hline
				& \multicolumn{4}{c|}{$\text{RPE}$} & \multicolumn{2}{c}{$\text{MCR}$} \\
				\cline{2-7}
				TensorNN-op & $ 0.0736 (0.0008) $ & $ \mathbf{0.1153} (0.0024) $ & $ \mathbf{0.0658} (0.0005) $ & $ \mathbf{0.0874} (0.0009) $ & $ \mathbf{0.1785} (0.0015) $ & $\mathbf{0.2229} (0.0020) $ \\
				CNN & $0.1335 (0.0016)$ & $0.1590 (0.0021)$ & $0.1296 (0.0008)$ & $0.1573 (0.0010)$ & $0.1890 (0.0017)$ & $0.2681 (0.0087)$ \\
				STAR & $0.0663 (0.0005)$ & $0.2410 (0.0019)$ & $0.0841 (0.0006)$ & $0.1571 (0.0010)$ & $0.1807 (0.0015)$ & $0.2444 (0.0021)$ \\
				TLR & $\mathbf{0.0653} (0.0005)$ & $0.1866 (0.0013)$ & $0.0855 (0.0006)$ & $0.1363 (0.0012)$ & $0.1883 (0.0014)$ & $0.2495 (0.0020)$ \\
				\hline
				\hline
				& \multicolumn{6}{c}{$\text{MEE}$} \\
				\cline{2-7}
				TensorNN-op & $ 0.0279 (0.0003) $ & $  \mathbf{0.0914} (0.0021) $ & $ \mathbf{0.0469} (0.0004) $ & $ \mathbf{0.0366} (0.0004) $ & $ 0.0694 (0.0019) $ & $\mathbf{0.1063} (0.0024) $ \\
				CNN & $0.0586 (0.0007)$ & $0.1310 (0.0016)$ & $0.1113 (0.0005)$ & $0.0747 (0.0005)$ & $0.1070 (0.0023)$ & $0.1356 (0.0060)$ \\
				STAR & $0.0241 (0.0001)$ & $0.2090 (0.0010)$ & $0.0653 (0.0004)$ & $0.0747 (0.0004)$ & $\mathbf{0.0633} (0.0009)$ & $0.1281 (0.0006)$ \\
				TLR & $\mathbf{0.0234} (0.0001)$ & $0.1576 (0.0008)$ & $0.0668 (0.0004)$ & $0.0636 (0.0004)$ & $0.0684 (0.0012)$ & $0.1286 (0.0004)$ \\
				\hline
		\end{tabular} }
		\label{table_simu1}
	\end{table}

	\section{Real Data Experiments}\label{sec:real}

	We consider four real  data from two types of  tensor data: fluorescence excitation-emission matrix (EEM) data and facial image data. For the fluorescence experiments, we use the sugar fluorescence data \citep{BRO1999133} and the synthetic fluorescence data \citep{RIU200335}. The sugar data consist of 268 samples, each represented by a \(571\times 7\) EEM,
	with ash content as the response. The synthetic fluorescence data contain hydroquinone measurements and, after preprocessing, include 26 samples with \(69\times 24\) EEM covariates.

	For the image experiments, we use the Labeled Faces in the Wild (LFW) dataset \citep{huang2008labeled} with facial attribute annotations \citep{kumar2009attribute}. After preprocessing, each image is represented as a \(45\times45\times3\) tensor. We use the smiling attribute as the response in a regression task and the mouth-closed attribute in a classification task. For classification, the mouth-closed attribute is dichotomized by its sign, \(y_i=1_{\{\mathrm{attr}_i>0\}}\). All tensor covariates are normalized to \([0,1]\).

	For the fluorescence datasets, experiments are repeated 100 times because of the limited sample sizes. For the facial datasets, experiments are repeated 50 times. In each replicate, the data are randomly split into training, validation, and test sets in the ratio $5:1:1$.

	\begin{table}[t]
		\centering
		\spacingset{1}
		\caption{
			Data size, tensor dimension, prediction performance, and selected feature numbers in the real-data experiments. Prediction performance is measured by average RPE for regression tasks and average MCR for the classification task.
			The last column reports the median selected feature number \(\widehat K\) for TensorNN-sc. Numbers in parentheses are standard deviations for RPE/MCR, and MADs for \(\widehat K\) over repeated random splits. The best prediction performance in each experiment is boldfaced.
		}
		\spacingset{1.4}
		\resizebox{\textwidth}{!}{
			\begin{tabular}{c|c| c | c c c  c c| c | c }
				\hline
				Real data & Data size \& dimension &  & TensorNN-op& TensorNN-sc & CNN & STAR & TLR & & TensorNN-sc\\
				\hline
				\hline
				Sugar& \multirow{2}{*}{$268$ \& $571\times 7$} &  \multirow{5}{*}{RPE}  &$ 0.1963 $& $ \mathbf{0.1899}$ & $ 0.2849 $& $ 0.1996 $& $ 0.2051 $ & \multirow{7}{*}{$\widehat K$} & $11.5$
				\vspace{-3mm}\\		
				Fluorescence& &   &$  (0.0064) $& $(0.0058)$ & $ (0.0098) $& $ (0.0051) $& $ (0.0057)$ & & $(2.5)$ \\		
				\cline{1-2} 	\cline{4-8}  \cline{10-10}
				Synthetic  & \multirow{2}{*}{$26$ \& $69\times 24$} &   & $ 0.1565$& $ \mathbf{0.1519}   $ & $ 0.1850 $& -- & $ 0.3626 $ & & $10.0$
				\vspace{-3mm}\\		
				Fluorescence &	&    & $ (0.0095) $& $(0.0101) $ & $(0.0111) $& -- & $  (0.0074)$ & & $(0.0)$   \\
				\cline{1-2} 	\cline{4-8} \cline{10-10}
				\multirow{2}{*}{Smiling}& \multirow{2}{*}{$5721$ \& $45\times 45\times 3$} &   & $ 0.1542  $& $ 0.1538$ & $ \mathbf{0.1446}$& $ 0.1869  $&	$0.1892  $ & & $46.0$
				\vspace{-3mm}\\		
				&  &   & $ (0.0013) $& $(0.0012)$ & $ (0.0015) $& $  (0.0014) $& $ (0.0016) $&  & $(12.0)$ \\
				\cline{1-8} 	\cline{10-10}
				\multirow{2}{*}{Mouth Closed} & \multirow{2}{*}{$5721$ \& $45\times 45\times 3$}& \multirow{2}{*}{MCR} & $ \mathbf{0.1549}  $& $0.1561 $ & $0.1657 $& $ 0.1974$& $ 0.1641	$ & &  $32.0$ \vspace{-3mm}\\
				& &  & $  (0.0016) $& $(0.0018) $ & $ (0.0018)$& $  (0.0021)$& $ (0.0021)$ & & $(8.0)$ \\		
				\hline
		\end{tabular}}
		\label{table_real}
	\end{table}

	Table~\ref{table_real} reports the prediction performance of all methods and the selected feature numbers for TensorNN-sc. Across the four real-data experiments, TensorNN achieves the best or near-best performance. For the sugar fluorescence data, TensorNN attains the lowest RPE, slightly improving over STAR and TLR while substantially outperforming CNN. For the synthetic fluorescence data, where the sample size is only 26, TensorNN again achieves the best performance, followed by CNN, whereas TLR performs much worse. STAR is not reported because it attains \(\mathrm{RPE}>1\), even underperforming the trivial zero predictor.
	These results suggest that the proposed method can adaptively approximate the regression function even with limited training data or highly unbalanced tensor dimensions.
	This robustness stems from TensorNN's ability to effectively learn low-dimensional  representations together with a compact outer neural network tailored to the selected  dimension, thereby reducing the effective number of parameters and, in turn, alleviating the method's reliance on large sample sizes.

	For the facial image data, CNN achieves the lowest RPE on the smiling prediction task, while TensorNN remains competitive and clearly outperforms STAR and TLR. This is reasonable because smiling is reflected by spatially distributed facial cues, and convolutional filters are effective at capturing  spatial patterns. For the mouth-closed classification task, however, TensorNN achieves the lowest MCR and outperforms CNN, STAR, and TLR. This task is more closely associated with a localized and structured mouth region, which can be effectively captured by the sparse CP-based tensor feature extractors. These results suggest that TensorNN is competitive for prediction with image tensor covariates and can be particularly advantageous when the predictive signal admits a low-dimensional structured representation.

	The selected feature numbers reported in Table~\ref{table_real} are also informative. For the two facial image tasks, TensorNN-sc selects more features for smiling prediction than for mouth-closed classification. This pattern is consistent with the nature of the two attributes: smiling involves coordinated changes across several facial regions, including the eyes, eyebrows, cheeks, and mouth, whereas the mouth-closed attribute is mainly associated with a localized mouth region.
	The magnitudes of $\widehat K$ also  suggest that rich facial details may be needed to characterize human expressions. Nevertheless, by encouraging model sparsity while retaining competitive predictive accuracy during model selection, we can see that the dominant features remain a minority: reducing the median number of features from $46$ to $9.5$ for the smiling attribute and from $32$ to $3.5$ for the mouth closed attribute only increases the relative error by $3.64\%$ and $1.60\%$, respectively.

	\begin{figure}[thbp]
		\centering
		\includegraphics[width = \textwidth]{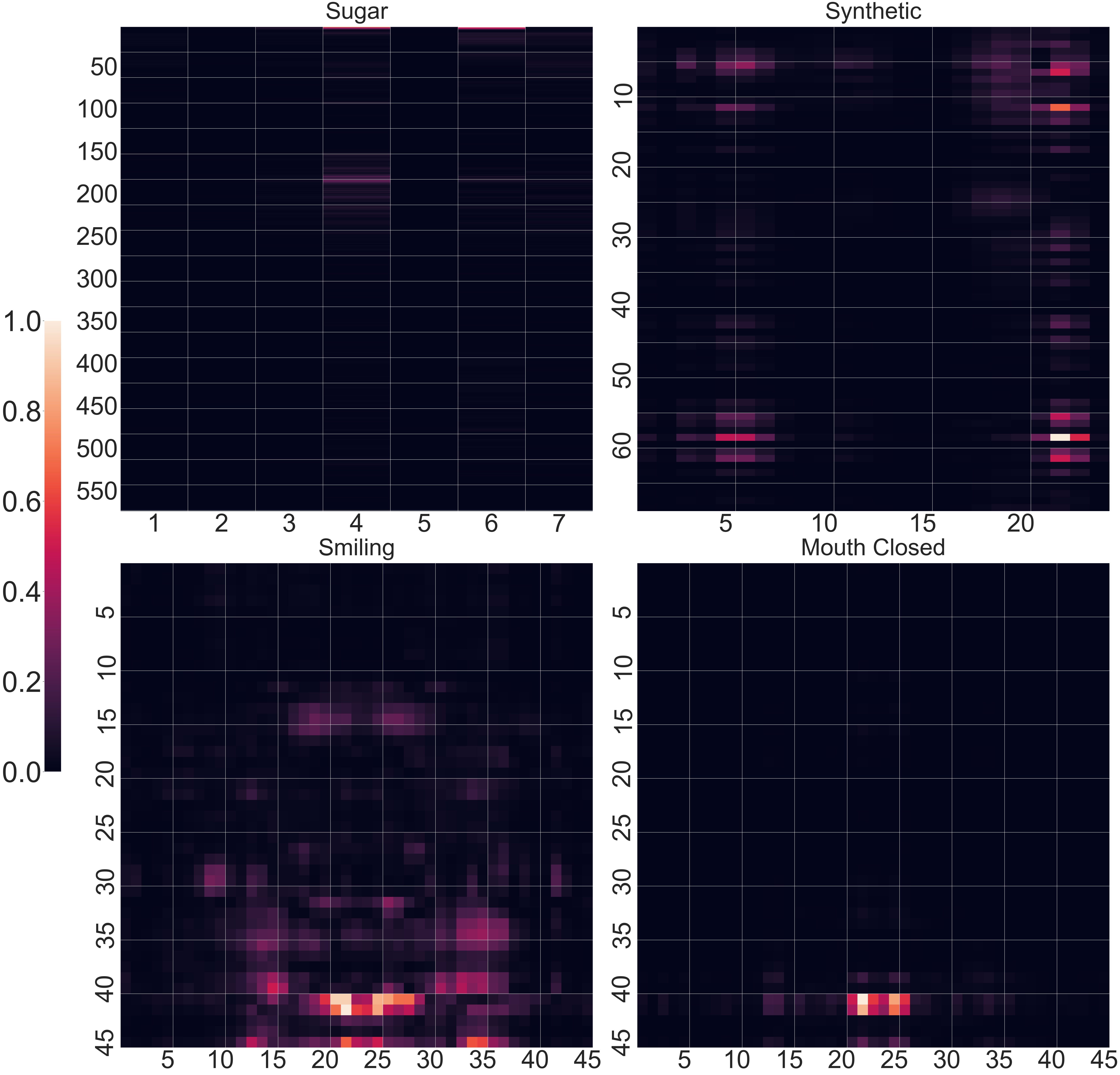}
		\spacingset{1}
		\caption{The heatmaps for all the real data experiments obtained by TensorNN-sc. Columns 1--4 correspond to the sugar fluorescence, synthetic fluorescence, smiling facial, and mouth closed facial experiments, respectively. The shade of square $(i,j)$ in the heatmaps represents the relative importance of the $(i,j)$-element of tensor covariate $\mX$.}
		\label{heatmap_real}
	\end{figure}

	We next examine the interpretability of TensorNN. As in Section~\ref{simu_result}, we visualize the estimated active regions through the aggregated importance tensor \(\widehat{\mV}^2\), which is constructed from the learned tensor feature extractors. For facial images, the importance values over the three RGB channels are summed to obtain a two-dimensional heatmap. Figure~\ref{heatmap_real} displays the resulting heatmaps for  TensorNN-sc, where brighter entries indicate tensor elements with larger estimated importance.
	For the sugar fluorescence data, TensorNN-sc highlights several important excitation wavelengths, with vertically structured patterns over emission wavelengths, suggesting that the learned features capture organized EEM signals rather than isolated tensor entries.
	For the synthetic fluorescence data, TensorNN-sc identifies sparse and localized EEM regions, mainly corresponding to emission wavelengths around 249--265, 293--413, and 455--469 nm, with excitation wavelengths around 215--225 and 285--315 nm.
	For the facial image data, the highlighted regions are consistent with the corresponding attributes: the smiling task involves broader facial areas, including the mouth, cheeks, and eye-related regions, whereas the mouth-closed task mainly concentrates around the mouth region. These results demonstrate that TensorNN provides intrinsic tensor-level interpretability while retaining strong predictive performance.

	\section{Discussion}\label{sec_discuss}

	This paper proposes a tensor neural network framework for learning  predictive and interpretable representations from tensor-valued covariates.
	The key idea is to identify a minimal set of informative tensor features that preserves the multiway structure of tensor covariates.
	By integrating tensor decomposition,  representation learning, and structured sparsity in both the feature and network components, the proposed method adaptively selects the representation dimension and provides a flexible, predictive, and interpretable alternative to classical tensor regression models and CNNs.

	The proposed three-part weighted penalty is key to achieving this adaptivity.
	The feature-level penalty removes redundant extractors, the component-level penalty induces local sparsity within each extractor, and the network penalty compresses unnecessary weights in the outer neural network.
	Consequently, one may start from a sufficiently rich working model with a relatively large feature dimension and network size, while the penalty adaptively controls the effective representation dimension and model complexity.
	The relative penalty weights derived from the theoretical error bound lead to the TensorNN-op implementation, which requires only lightweight tuning of the overall regularization strength and a rough choice of network depth.
	When feature recovery is the primary goal, TensorNN-sc further tunes the feature-level weight \(\tau_1\), as suggested by the selection-consistency theory.

	Theoretically,
	the oracle-type prediction bound in Theorem~\ref{converge_thm1} shows that the leading error rate is governed by the minimal feature dimension \(K^*\), rather than by the ambient tensor dimension \(\mathcal{D}\).
	Under the stated regularity conditions, TensorNN attains the \(K^*\)-dimensional nonparametric rate \(n^{-2\xi/(2\xi+K^*)}\), up to logarithmic factors, even though estimation is carried out over an enlarged working model.
	Theorem~\ref{selection_thm2} shows that, under an additional feature signal-strength condition, increasing the feature-level penalty yields consistent recovery of \(K^*\) without sacrificing the oracle prediction rate.

	The  results in Tables~\ref{table_simu}, \ref{table_simu1}, and \ref{table_real} broadly support these theoretical conclusions.
	Across both simulations and real-data applications, TensorNN achieves stronger predictive performance than the competing methods while producing interpretable low-dimensional tensor representations.
	Its advantage is most evident when the underlying model is complex.
	Overall, the empirical evidence suggests that TensorNN can adapt to different types of tensor covariates and signal structures.

	Unlike CNNs, whose interpretation typically relies on post-hoc tools, TensorNN provides model-level active-region summaries through its learned tensor feature extractors.
	The simulation and real-data heatmaps in Figures~\ref{heatmap_simu} and \ref{heatmap_real} show that TensorNN can identify meaningful tensor regions associated with the response.
	This built-in interpretability is particularly valuable for non-image tensor applications, where generic convolutional architectures and post-hoc visualization tools may not yield stable or scientifically meaningful explanations.

	Two limitations remain.
	First, the current framework uses general tensor feature extractors; incorporating additional domain-specific structure may further improve performance and interpretation.
	For example, spatial smoothness can be imposed for image tensors, while temporal or contextual dependence can be incorporated by combining tensor feature extraction with recurrent or Transformer-based architectures.
	Second, the active-region heatmaps should be viewed as model-based summaries rather than  causal explanations.
	These limitations provide promising directions for extending TensorNN to broader tensor learning problems

	\spacingset{0.9}

\end{document}